\documentclass[
    ,final            
,numberedheadings 
  ]
  {aipproc}

\layoutstyle{6x9}

\begin{document}

\graphicspath{{/afs/ifh.de/user/n/nowakw/HERMES/TALKS/EPS/}}

\title{The Angular Momentum Structure \\ of the Nucleon}


\author{Wolf-Dieter Nowak}{
  address={DESY, D-15738 Zeuthen, Germany}
}


\begin{abstract}
The proton spin budget is discussed. Results are presented from inclusive 
and semi-inclusive deep inelastic scattering and from deeply virtual Compton 
scattering. They permit interpretations
towards the determination of various contributions to the proton spin. 
\end{abstract}

\maketitle



\section{Introduction}


One of the most fundamental questions when studying the structure of the
nucleon is how its spin $\frac{1}{2}$ is decomposed into the spins and 
orbital angular momenta of quarks and gluons. A milestone in the history 
of this field was the appearance in the late 1980s of an unexpected
result of the European 
Muon Collaboration (EMC)~\cite{Ashman:1987hv,Ashman:1989ig}, which 
aroused great interest. Experimental advances in polarized-target 
technology in conjunction with the high degree of polarization of 
the ${\cal{O}}(100)$ GeV polarized muon 
beam at the {\sc{Cern}} SPS, had made it possible to measure the 
double-spin asymmetry $A_{LL}$ in the cross section of inclusive
Deeply Inelastic Scattering (DIS) of leptons on nuclei. From this 
asymmetry the quark contribution $\Delta \Sigma$ to the nucleon spin 
was determined. It was found to be much smaller than the predictions 
within the widely accepted picture of three valence quarks constituting 
essential components of the nucleon. This finding was hence called a
`spin-puzzle'. Several new experiments were proposed and 
a wealth of theoretical studies initiated to solve it. Among the latter 
was another milestone, namely the finding~\cite{Ji:1996ek} by Ji that 
the total angular momentum $J_f$ of quark species $f$ and that of gluons 
($J_G$) can be experimentally accessed in principle, by evaluating second 
moments of Generalized Parton Distributions (GPDs)~\cite{Ji:1996nm}
which had already been discussed 10 years 
before by M\"uller et al.~\cite{Mueller:1998fv}. GPDs were more recently
interpreted by Burkardt~\cite{Burkardt:2000za} to describe simultaneously
the transverse localization $b_\perp$ of a parton in the nucleon
at a given fraction $x$ the parton carries of the longitudinal 
momentum $P$ of the nucleon. This description is sometimes also 
refered to as 3-dimensional picture of the nucleon.

In relativistic quantum mechanics, the spin of the nucleon can be 
decomposed into the total angular momenta of quarks and gluons. The `Ji 
representation'~\cite{Ji:1996nm} of this nucleon spin sum rule can be 
written in a covariant way as:
\begin{eqnarray}
\frac{1}{2} = J_Q(\mu^2) + J_G(\mu^2),
\label{Eq:SpinBudget1}
\end{eqnarray}
with $J_Q \equiv \sum_{f} J_f$. In the context of Quantum 
Chromodynamics (QCD), the gauge theory of strong interactions, the 
individual terms in any decomposition of the nucleon spin depend on the 
factorization scheme and on the renormalization scale $\mu^2$. Results 
in different schemes are uniquely related to one another; mostly they 
are given in the so-called 
\( \overline{\rm MS} \) scheme. The hard scale $\mu^2$ of the interaction 
can be chosen freely and independently of experiment. In lowest order 
perturbation theory (pQCD), lepton-nucleon scattering is treated in the 
one-photon-exchange approximation and $\mu^2$ is then identified with the 
virtuality of the exchanged virtual photon, $Q^2 \equiv -q^2$, with $q$ 
being its 4-momentum.
The value of $Q^2$ is commonly chosen to be at least a few GeV$^2$, 
a choice that is essential to keep all sorts of corrections as small 
as possible (most of them are suppressed by inverse powers of $Q$)  
when interpreting the data in terms of pQCD. Once an observable is known 
at one scale $Q^2_0$, QCD evolution equations~\cite{Altarelli:1977zs} can 
be used to calculate it at another value $Q^2_1$. We note that no 
practical way is known to determine the gluon total angular momentum $J_G$.
Still, it can be calculated as the difference $\frac{1}{2} - \sum_{f} J_f$.

With $\Delta\!\!\!\stackrel{(-)}{q}\!\!\!_f$ being the contribution from 
(anti)quark species $f$, the quark total angular momentum $J_Q$ in
(\ref{Eq:SpinBudget1}) can be replaced by the sum of the intrinsic quark spin 
contribution $\frac{1}{2} \Delta \Sigma \equiv \frac{1}{2} \sum_{f} 
\left( \Delta q_f + \Delta \bar{q}_f \right)$ plus the quark orbital 
angular momentum contribution $L_{Q} \equiv \sum_{f} L_f$. Taking the $z$ 
(infinite momentum) axis as the quantization axis, this leads to:
\begin{eqnarray}
\frac{1}{2} = S_z = \frac{1}{2}\Delta \Sigma + L_Q + J_G.
\label{Eq:SpinBudget2}
\end{eqnarray}
Here, only $\Delta \Sigma$ has a probabilistic 
interpretation as a parton number density. 

Recent experimental results on $\Delta \Sigma$ from inclusive DIS will
be presented below, followed by a description of the first 
steps towards a measurement of $J_u$ and $J_d$ in Deeply Virtual 
Compton Scattering (DVCS). Presently, no practical way is known to 
measure quark orbital angular momenta $L_f$ separately or their sum
$L_Q \equiv \sum_{f} L_f$ as a whole. Hence the quark spin budget 
equation
\begin{eqnarray}
J_Q = \frac{1}{2} \Delta \Sigma + L_Q 
\label{Eq:SpinBudget3}
\end{eqnarray}
cannot be tested yet. Still, once $J_u$ and $J_d$ will be known,
the valence quark orbital angular momentum $L_{u+d}$  can be obtained 
as the difference using (\ref{Eq:SpinBudget3}). 

Later in this article first results will be shown on $\Delta G$, the 
intrinsic gluon contribution to the nucleon spin. Then, also the 
difference $L_G = J_G - \Delta G$ could be calculated in 
principle~\cite{Ji:1996ek}.
However, while all observables discussed above are expectation values of 
gauge-invariant operators, for this difference no operator representation 
with any intrinsic relation to orbital angular momentum was identified yet, 
so that $L_G$ may remain a sterile definition (see 
\cite{Burkardt:2008jw,Burkardt:2008ua} for more 
detailed discussions). We note that both $L_Q$ and $L_G$ can be determined
in QCD calculations on the Lattice (see, e.g., \cite{Hagler:2007xi}).

Already back in 1989, as one of the first theoretical attempts to solve the 
spin puzzle, the alternative `Jaffe-Manohar 
representation'~\cite{Jaffe:1989jz} of the nucleon spin sum rule had been 
proposed, which is defined in light-cone ($A^+ = 0$) or DIS gauge. It 
represents a decomposition of the $z$ projection of the nucleon spin:
\begin{eqnarray}
\frac{1}{2} = S_z = \frac{1}{2}\Delta \Sigma + \Delta G + {\cal{L}},
\label{Eq:SpinBudget4}
\end{eqnarray}
where $\cal{L}$ is the parton (quark plus gluon) contribution to
the nucleon spin, which can even be further decomposed into quark and
gluon components: ${\cal{L}} = \sum_{f}{\cal{L}}_f + {\cal{L}}_G$.
In decomposition (\ref{Eq:SpinBudget4}) the (measurable) intrinsic gluon 
contribution $\Delta G$ to the nucleon spin appears explicitly and all 
terms have a probabilistic 
interpretation as a parton number density. No way to measure any of the
orbital angular momenta ${\cal{L}}, {\cal{L}}_f$, or ${\cal{L}}_G$ is
presently known nor can they be calculated in Lattice QCD. Still, 
${\cal{L}}$ can be determined as the difference using (\ref{Eq:SpinBudget4}). 

Comparing the Ji representation (\ref{Eq:SpinBudget2}) and the Jaffe-Manohar 
representation (\ref{Eq:SpinBudget4}) of the nucleon spin sum rule 
(\ref{Eq:SpinBudget1}) it can be seen that they have only $\Delta \Sigma$ in 
common, while all other components appear only in one or the other 
representation. The incommensurability of the two representations is 
discussed in \cite{Burkardt:2008jw,Burkardt:2008ua} by comparing the
representations of the operators, the expectation values of which appear
in one of the above discussed decompositions of the nucleon spin. 
It is concluded that, while both representations have their respective 
merits, according to present understanding and knowledge the components 
of the two representations must not be mixed because of incommensurate 
underlying operator representations.


\section{Results from Deeply Inelastic Scattering}

\begin{figure}[htb]  \label{Fig:Hermes2007g1d}
  \includegraphics[width=.4\textwidth]{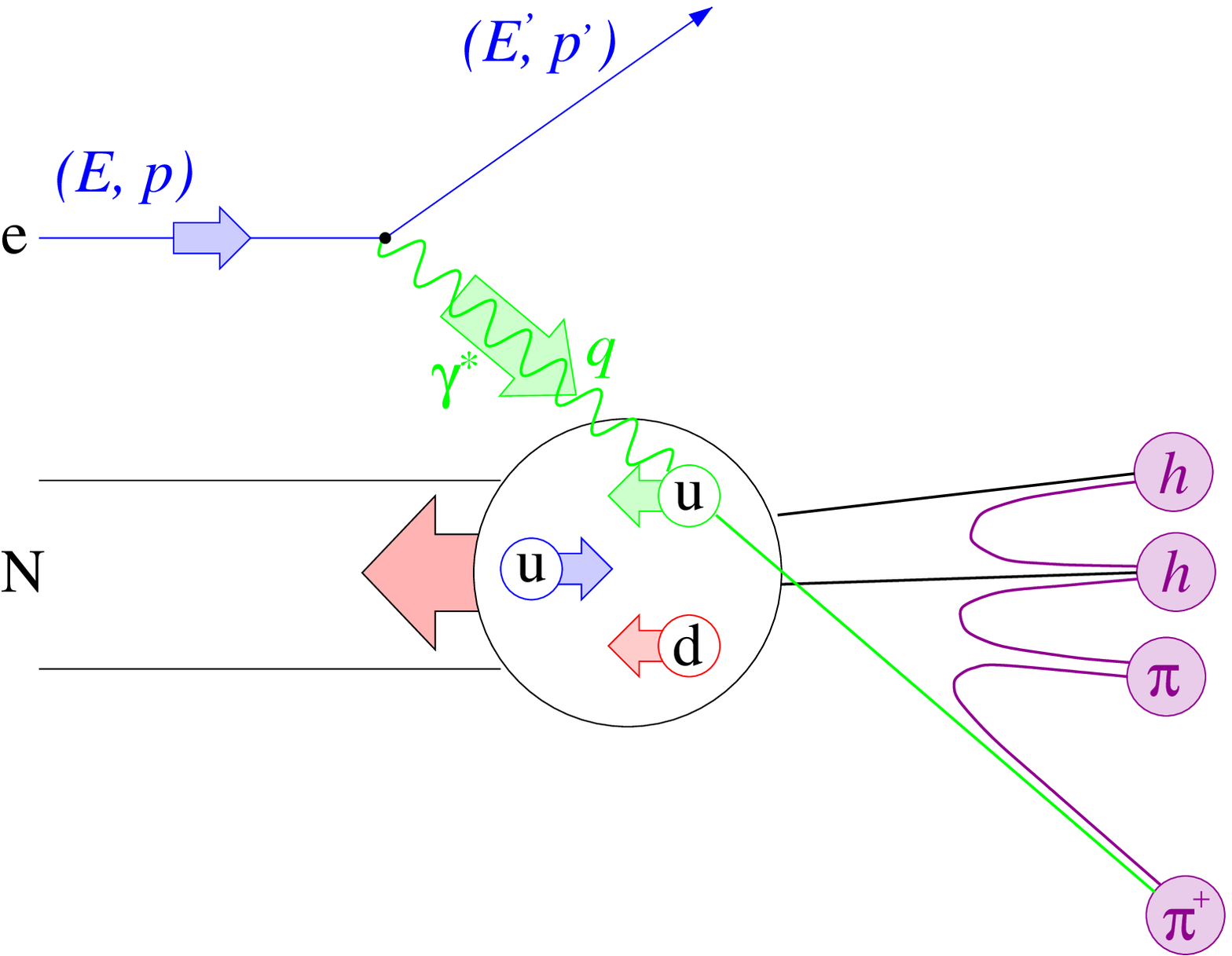}
  \hspace*{2mm}
  \includegraphics[width=.5\textwidth]{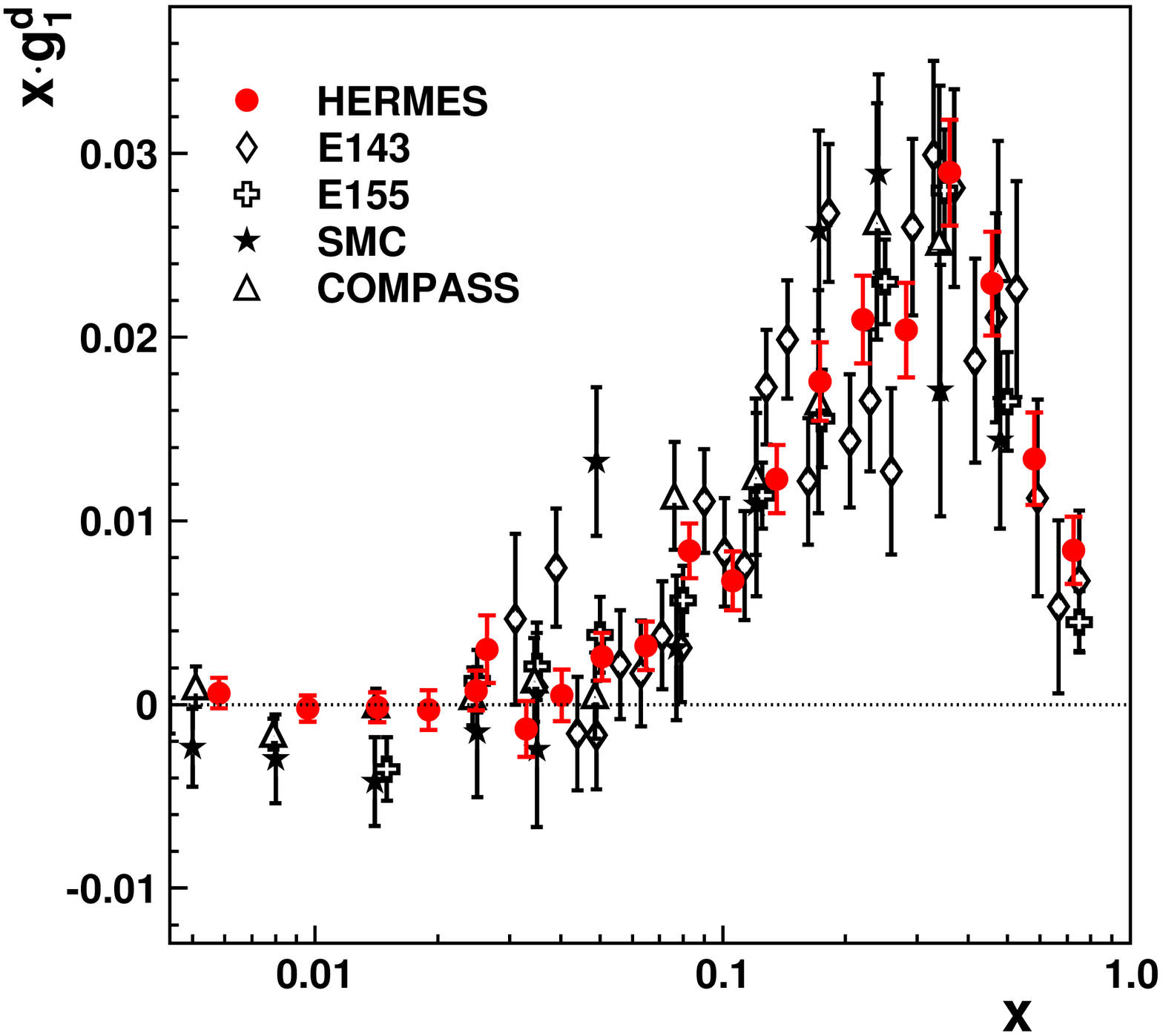}
  \caption{Left: Simplified schematic diagram of semi-inclusive DIS. 
Right: $x$ dependence of $x \cdot g_1^d$, as measured by 
{\sc{Hermes}}~\cite{Airapetian:2007mh}, E143~\cite{Abe:1998wq}, 
E155~\cite{Anthony:2000fn}, {\sc{SMC}}~\cite{Adeva:1999pa} and 
{\sc{Compass}}~\cite{Ageev:2005gh}. The figure is taken from 
\cite{Airapetian:2007mh}.}
\end{figure}

Doubly-polarized deeply inelastic scattering is sketched in the left 
panel of Fig.~\ref{Fig:Hermes2007g1d}. For inclusive DIS, in which 
only the scattered lepton in analyzed, the double-helicity asymmetry 
$A_{LL}$ is given by the DIS cross-section helicity difference,
$\sigma_{LL} \equiv \frac{1}{2} 
                    (\sigma^{\stackrel{\rightarrow}{\Leftarrow}} 
                   - \sigma^{\stackrel{\rightarrow}{\Rightarrow}})$,
normalized by
$\sigma_{UU} \equiv \frac{1}{2} 
                    (\sigma^{\stackrel{\rightarrow}{\Leftarrow}} 
                   + \sigma^{\stackrel{\rightarrow}{\Rightarrow}})$,
the unpolarized DIS cross section:
\begin{equation}
A_{LL} = \frac{\sigma_{LL}}{\sigma_{UU}} \propto  \frac{g_1}{F_1}.
\label{Eq:Aparallel}
\end{equation}
Here `U' and `L' stand for unpolarized and longitudinally polarized
(beam and target), and the single and double arrows denote the relative 
orientation of the spins of initial lepton and nucleon, respectively. 
As shown in (\ref{Eq:Aparallel}), this asymmetry can be approximated by 
the ratio of $g_1$ and $F_1$, the spin-dependent and spin-averaged 
structure functions of the nucleon, respectively. They can both be 
expressed in terms of respective quark distribution functions.

The most recent result on $g_1$ was obtained by the 
{\sc{Hermes}} collaboration~\cite{Airapetian:2007mh}. In the right panel 
of Fig.~\ref{Fig:Hermes2007g1d} the $x$ dependence of $x \cdot g_1^d$ 
is shown together with earlier results. In DIS, $x$ is identified with 
the Bjorken variable $x_B \equiv Q^2/(2P \cdot q)$. Using a few plausible 
assumptions, e.g. no significant contribution from the (unmeasured) 
small-$x_B$ region, the {\sc{Hermes}} data set was used to accomplish the 
presently most precise experimental determination of the quark 
contribution to the nucleon spin~\cite{Airapetian:2007mh}:   
\begin{equation}
\Delta \Sigma = 0.330 \pm 0.011_{theor.} \pm 0.025_{exp.} \pm 0.028_{evol.},
\label{Eq:Hermes2007g1d}
\end{equation}
where `exp.' includes statistical, experimental systematic and 
parameterization uncertainties, while the `evol.' uncertainty arises from 
the necessity to evolve data to a common value of $Q^2$, here 5 GeV$^2$.

There exists no result of comparable accuracy yet on $\Delta G$, the 
intrinsic gluon contribution to the nucleon spin, mainly because in 
inclusive DIS the virtual photon probes only quarks in the nucleon. The 
sensitivity to gluons in the nucleon can be enhanced when measuring hadrons 
in the final state, as those may be produced through additional QCD 
subprocesses, in particular photon-gluon fusion. Pioneering results 
were obtained in `direct' measurements of the gluon polarization 
$\frac{\Delta G}{G}(x)$ in hadron photoproduction at very low $Q^2$,
although only in the limited $x$ range covered by the data. Here,
not only photon-gluon but also quark-gluon collisions occur because 
the hadronic content of the virtual photon gives rise to 
`resolved-photon' subprocesses. Therefore, the resulting complicated 
mixture of subprocesses requires models in Monte Carlo simulations 
to describe the fractional contributions of the subprocesses within 
the experimental acceptance,
so that such analyses were only performed in leading order and, moreover,
the final result always remains model dependent. At these low values of
$Q^2$ the hard scale of the subprocess is obtained by either selecting
large transverse momenta of produced hadrons or by measuring charmed 
hadron production where the mass of the charm quark gives the hard scale.
 
The two presently most precise leading-order determinations of
$\frac{\Delta G}{G}(x)$ were accomplished by 
{\sc{Compass}}~\cite{Ageev:2005gh} analysing
high-$p_t$ hadron pairs produced for $Q^2 < 1$ GeV$^2$:
\begin{equation}  \label{DeltaG_Compass}
\frac{\Delta G}{G}|_{\langle x \rangle \simeq 0.085} = 
                  0.016 \pm 0.058_{stat} \pm 0.055_{syst},
\end{equation}
and by {\sc{Hermes}}~\cite{Liebing:2007bx} analysing
high-$p_t$ single hadrons from quasi-real photoproduction:
\begin{equation}  \label{DeltaG_Hermes}
\frac{\Delta G}{G}|_{\langle x \rangle \simeq 0.22} = 
                  0.071 \pm 0.034_{stat} \pm 0.010_{sys-exp}
\pm^{0.127}_{0.105~sys-Models}.
\end{equation}

In an alternative approach, sensitivity to the gluon content of 
the nucleon is found by fitting the $Q^2$ evolution 
of the moments of $g_1$ to the data, as this 
evolution involves gluon radiation by quarks and quark pair 
production by gluons. For such QCD fits none of the above 
described complications in direct measurements apply. 
Nevertheless, results on $\Delta G$ from earlier QCD analyses 
of only inclusive data showed quite large uncertainties, 
although predefined functional shapes were fitted and boundary 
conditions were imposed. The sensitivity to the gluon 
polarization in the nucleon and hence the accuracy of the fit 
can be improved by including data from semi-inclusive 
DIS (SIDIS) and doubly-polarized $pp$ collisions. In a very 
recent next-to-leading order QCD analysis~\cite{florian:2007hc}, 
(semi-)inclusive data on $A_{LL}^{p,d}$ from {\sc{Compass, JLab}} 
and {\sc{Hermes}} were combined with {\sc{Phenix}} and {\sc{Star}} 
$pp$ data on $A_{LL}^{\pi^0}$. Several results on $\Delta G$ are 
given at $Q^2$ = 10 GeV$^2$ for various ranges in $x$ and also 
for various ways to estimate the uncertainties. For the purpose 
of this article it appears sufficient to interpret their result 
as follows:
\begin{equation}
\Delta G \approx - 0.1 \pm 0.1.
\label{Eq:DeltaG_Result}
\end{equation}
The results (\ref{DeltaG_Compass}) and (\ref{DeltaG_Hermes}) from 
the leading-order direct determinations of the gluon polarization 
are consistent with the fit result (\ref{Eq:DeltaG_Result}) on 
$\Delta G$. More precise results on the gluon contribution to the 
nucleon spin can be expected when more data from doubly-polarized 
$pp$ collisions at {\sc{Rhic}} will become available.

Combining the DIS results (\ref{Eq:Hermes2007g1d}) and 
(\ref{Eq:DeltaG_Result}), the Jaffe-Manohar representation
(\ref{Eq:SpinBudget4}) of the nucleon spin sum rule allows the
determination of the parton orbital angular momentum in this 
representation:
\begin{equation}
{\cal{L}} \approx 0.4 \pm 0.1.
\end{equation}


\section{Results from Deeply Virtual Compton Scattering }

\begin{figure}[htb]  \label{fig:dvcs_bh_angles}
  \includegraphics[width=.22\textwidth,angle=270]{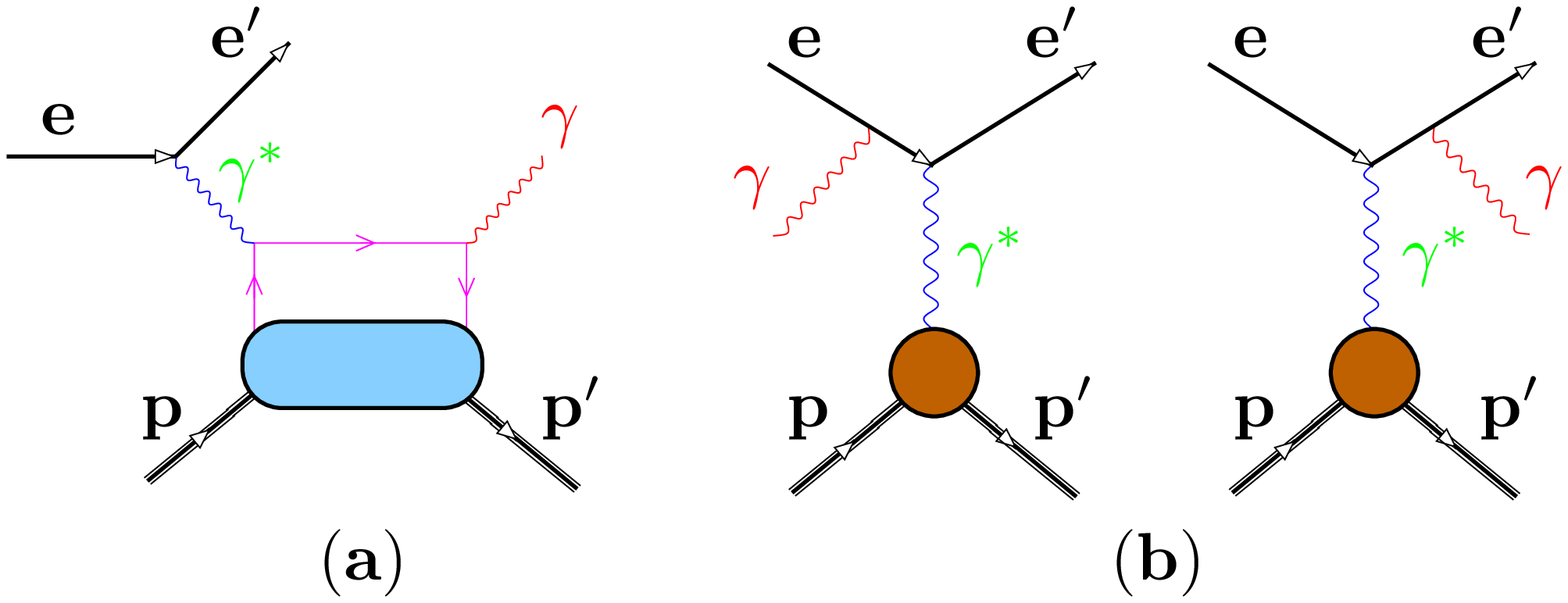}
  \hspace*{4mm}
  \includegraphics[width=.2\textwidth,angle=270]{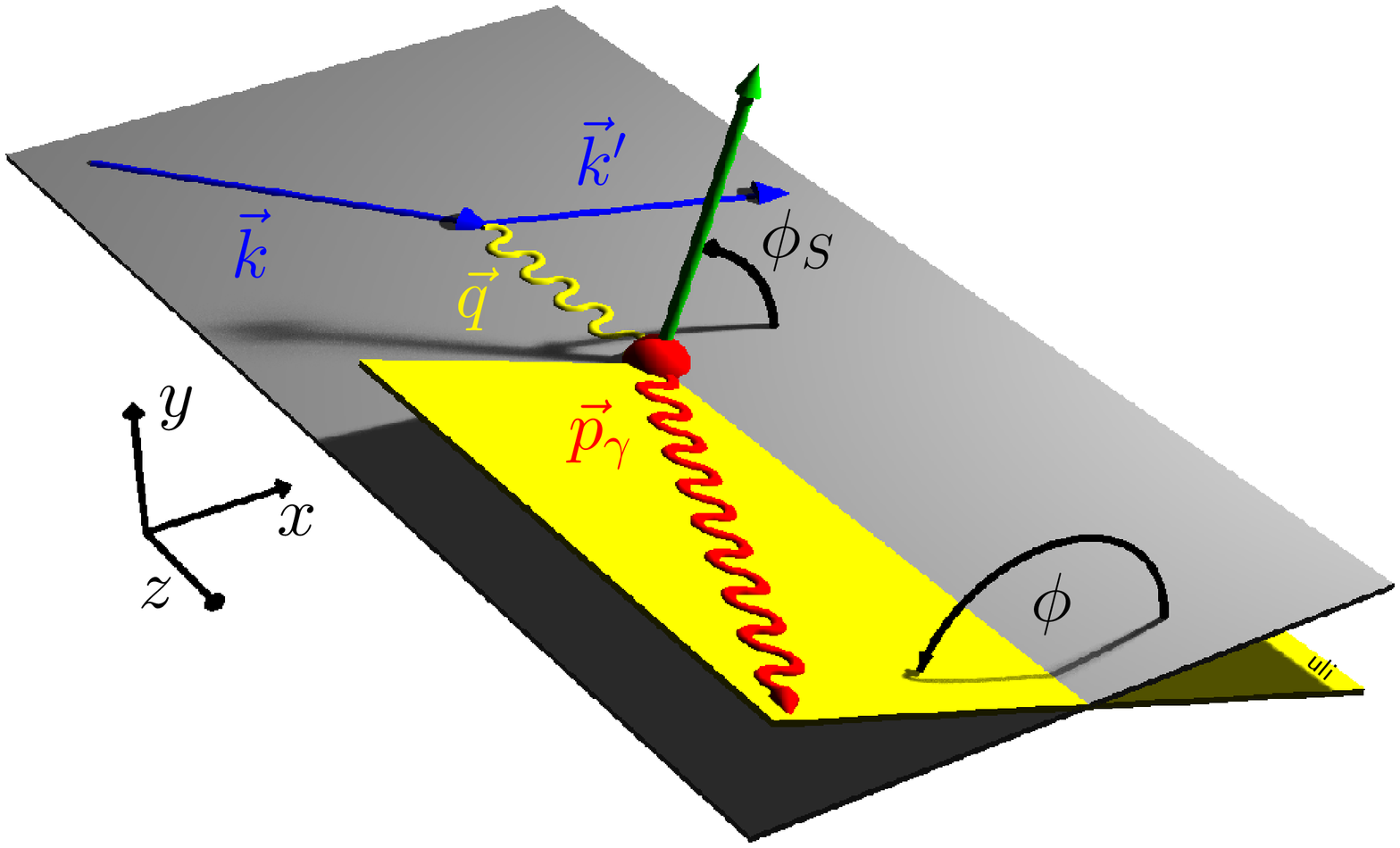}
  \caption{Left (a): `Handbag' diagram in DVCS.
  Middle (b): Bethe-Heitler process. Right: Definition of azimuthal angles 
of produced real photon ($\phi$) and target spin vector ($\phi_S$).}
\end{figure}

Exclusive production of a real photon from a virtual one, as
illustrated in Fig.~\ref{fig:dvcs_bh_angles}(a), is called
Deeply Virtual Compton Scattering (DVCS). The competing 
Bethe-Heitler (BH) process, illustrated in 
Fig.~\ref{fig:dvcs_bh_angles}(b), has an identical final state 
so that the two processes interfere on the level of amplitudes:
\begin{equation}
\frac{d\sigma(\mathrm{eN\rightarrow eN\gamma})}
{dx_BdQ^2d|t|d\phi}
           \propto \left|\mathcal{T}_{BH}\right|^2+
           \left|\mathcal{T}_{DVCS}\right|^2+
           \underbrace{\mathcal{T}_{BH}\mathcal{T}_{DVCS}^*
           +\mathcal{T}_{BH}^*\mathcal{T}_{DVCS}.}_{I}
\label{Eq:EprodXsect}
\end{equation}
Here $\phi$ is the azimuthal angle between the scattering plane, 
spanned by the incoming and outgoing leptons, and the production 
plane spanned by the virtual photon and the produced real photon
(see right panel of Fig.~\ref{fig:dvcs_bh_angles}). The 
Mandelstam momentum transfer between initial and final nucleon 
is denoted by $t$.
The BH amplitude $\mathcal{T}_{BH}$ is exactly calculable in leading-order
QED using the knowledge of the elastic nucleon Dirac and Pauli form 
factors $F_1$ and $F_2$. The DVCS contribution 
$\left|\mathcal{T}_{DVCS}\right|^2$ can be obtained by integrating over 
the azimuthal dependence of the cross section. It is sizeable 
at collider energies and on the few percent level at fixed-target energies, 
while the opposite holds for the contribution of the interference term I.

This interference term is of interest, as the measurement of its
azimuthal dependence opens access to the complex-valued DVCS process
amplitude. In this way both $\Re{\rm (I)}$ and $\Im{\rm (I)}$ become 
accessible~\cite{Diehl:1997bu}. Using a $1/Q$ expansion of the 
electroproduction cross section (\ref{Eq:EprodXsect}) the interference 
term can be represented in lowest order as a sum of azimuthal 
harmonics~\cite{Belitsky:2001ns}. Here, besides important kinematic 
factors, the coefficients of the $\cos$ ($\sin$) terms are real 
(imaginary) parts of certain Compton helicity amplitudes describing 
the DVCS process. In the interpretation of experimental data, only
the nucleon-helicity-conserving twist-2 amplitude is considered,
which can be written as a linear combination of $F_1$ and $F_2$ 
with Compton Form Factors (CFFs)
$\mathcal{F} (\mathcal{F} = \mathcal{H,E,\widetilde{H},\widetilde{E}})$. 
These CFFs are flavour sums of convolutions of the respective GPDs 
$H^f,E^f,\widetilde{H}^f,\widetilde{E}^f$ with hard-scattering kernels. 
In such a convolution the GPD dependence on the momentum fraction $x$ is 
integrated out, so that -- in contrast to DIS -- DVCS does not permit a
direct determination of the $x$ dependence of GPDs. Besides on $Q^2$ and 
$x$, GPDs also depend on $\xi$ and $t$ with the longitudinal momentum 
fraction $\xi$ describing the `skewness' of the handbag diagram shown in 
Fig.~\ref{fig:dvcs_bh_angles}(a). In the forward limit $\xi=0$, the quark 
taken out from the nucleon and the one put back into it carry identical 
momentum fractions and the GPDs $H^f$ and $\widetilde{H}^f$ become the 
spin-averaged and spin-dependent PDFs $q_f$ and $\Delta q_f$, respectively.

Various cross-section differences (or asymmetries) can be measured with 
respect to either beam spin or charge, or target polarization.
They filter out certain parts of the interference term, i.e.,
they are represented by certain linear combinations of CFFs.
Once the $t$-dependence of the GPDs $H^f$ and $E^f$ 
will have been determined, the second moment of their sum 
can be used to obtain the total angular momentum of quarks 
in the nucleon through the Ji sum rule~\cite{Ji:1996ek}:
\begin{equation}
J_f = \frac{1}{2} \lim_{t \rightarrow 0} \int\!\! {\rm d}x\,x
\left[H^f(x,\xi,t)+E^f(x,\xi,t)\right].
\label{eq:Ji_SumRule}
\end{equation}

\begin{figure}[htb]  \label{HermesBCA}
  \includegraphics[width=.67\textwidth]{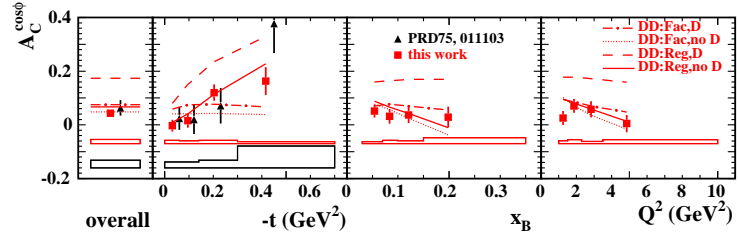}
  \caption{
Azimuthal amplitude describing the dependence of the 
interference term on the beam charge ($A_C$). The triangles (shifted 
right for visibility) represent previous results~\cite{Airapetian:2006zr}, 
while most recent data~\cite{:2008jga} are represented by squares. The 
error bars (bands) represent the statistical (systematic) uncertainties. 
The curves labelled 'DD' are calculations of variants of a 
double-distribution GPD model~\cite{Vanderhaeghen:1999xj,Goeke:2001tz} using 
$b_v=\infty$ and $b_s=1$ as profile parameters for valence and sea quarks.
This figure shows the top panel of figure 4 of \cite{:2008jga}.}
\end{figure}

\begin{figure}[htb]  \label{HermesTTSA}
  \includegraphics[width=.67\textwidth]{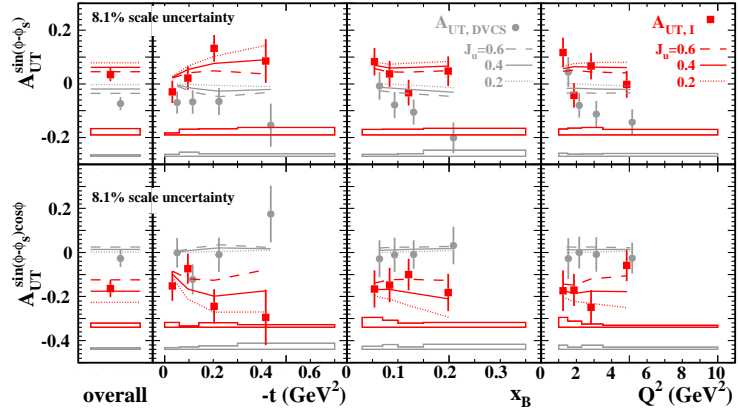}
  \caption{
Asymmetry amplitudes describing the dependence of the 
squared DVCS amplitude (circles, $A_{UT,DVCS}$) and the interference term 
(squares, $A_{UT,I}$) on the transverse target polarisation.
The circles (squares) are shifted right (left) for visibility. 
The error bars represent the statistical uncertainties, while the top 
(bottom) bands denote the systematic uncertainties for $A_{UT,I}$ 
($A_{UT,DVCS}$), excluding an 8.1~\% scale uncertainty from the target 
polarisation measurement. 
The curves are calculations of the GPD model variant 
(Reg, no D) shown in the previous figure as a continuous curve, with 
three different values for the $u$-quark total angular momentum $J_u$ 
and fixed $d$-quark total angular momentum $J_d=0$~\cite{Ellinghaus:2005uc}.
This figure shows the first two panels of figure 5 of \cite{:2008jga}.}
\end{figure}
 
Pioneering DVCS data from the {\sc{Hermes}} experiment were 
analyzed~\cite{:2008jga}
to determine cross section asymmetries with respect to beam spin and 
charge on one hand, and with respect to beam charge and target 
polarization on the other. In each case all relevant `effective 
asymmetry amplitudes' were determined, which are the above mentioned 
coefficients of the $1/Q$ expansion in azimuthal harmonics (for a 
detailed description see \cite{:2008jga}). As an example, results
from a `combined' fit of the asymmetries with respect to beam charge
and transverse target polarization are shown in Figs.~\ref{HermesBCA}
and \ref{HermesTTSA}, respectively. The data in Fig.\ref{HermesTTSA} 
are compared to calculations using various GPD models, see caption.
In such models, the $u$ and $d$-quark total angular momenta $J_u$ and 
$J_d$ can be used as free parameters to parameterize the GPDs $E^u$ 
and $E^d$, respectively (see, e.g., \cite{Ellinghaus:2005uc}). The 
asymmetry amplitudes calculated from the GPD model of 
\cite{Vanderhaeghen:1999xj} were fitted to the 
{\sc{Hermes}} data with transverse target polarization, resulting in 
the model-dependent constraint~\cite{:2008jga}:
\begin{equation}
\label{eq:constr}
J_u+J_d/2.8=0.49\pm0.17(\mathrm{exp_{tot}}).
\end{equation}
This result shows that the data are already able to yield
information on the total angular momenta of $u$ and $d$-quarks, while
the limitations of the GPD model used prevent clear conclusions.

In leading-order approximation the values of the relevant (combination 
of) GPD(s) along the `cross-over trajectory' $x=\xi$ contain the 
physical content of GPDs that is accessible in DVCS, while their 
{\em shape} in the ($x, \xi$) plane is not accessible. In this context, 
predefined ($x,\xi$) dependences used in available GPD models may be 
too inflexible and even ambigous~\cite{Kumericki:2008di}. At fixed $t$, 
CFFs satisfy dispersion 
relations~\cite{Frankfurt:1997ha,Teryaev:2005uj,Kumericki:2008di} 
that allow the evaluation of the real part of a CFF from its imaginary part. 
By combining dispersion relation and operator 
expansion techniques convergence was proved for the conformal partial 
wave expansion of the VCS amplitude~\cite{Kumericki:2007sa} and this
new formalism was successfully applied to fitting DVCS observables in
NNLO at low $x_B$. No global fits to all existing DVCS data exist yet. 

The information on the $u$ and $d$-quark total angular momenta $J_u$ 
and $J_d$ that is presently available from DVCS experiments appears 
neither reliable enough to determine the gluon total angular momentum 
$J_G$ through (\ref{Eq:SpinBudget1}) nor through (\ref{Eq:SpinBudget3}) 
the valence quark orbital angular momentum $L_{u+d}$ in the Ji 
representation.


\begin{theacknowledgments}
I'm deeply indebted to M. Burkardt \& A. Miller for very exciting, fruitful 
and long discussions towards paper \cite{Burkardt:2008jw} without which the 
first section 
of this article would not have been possible. I'm grateful to A. Miller 
for reading this manuscript.
\end{theacknowledgments}


\bibliographystyle{aipprocl} 


\bibliography{Trieste2008}




\end{document}